\documentclass[draft]{agujournal2019}
\usepackage{url} 
\usepackage{lineno}
\usepackage[finalnew]{trackchanges} 
\usepackage{soul}
\usepackage[T1]{fontenc} 
\usepackage{amssymb}
\newcommand{\Ex}{E$_\textrm{x}$}
\newcommand{\Ey}{E$_\textrm{y}$}

\newcommand{\Epred}[1]{E$_{\textrm{#1},pred}$}
\newcommand{\Epers}[1]{E$_{\textrm{#1},pers}$}

\newcommand{\GICsum}[1]{GIC#1$_{sum1h}$}
\newcommand{\GICabs}[1]{$\mid$GIC#1$\mid$}

\newcommand{\GICpredE}[1]{GIC#1$_{pred,E}$}
\newcommand{\GICpred}[1]{GIC#1$_{pred}$}
\newcommand{\GICpers}[1]{GIC#1$_{pers}$}

\draftfalse

\journalname{Space Weather}

\begin{document}


\title{Forecasting GICs and geoelectric fields from solar wind data using LSTMs: application in Austria}

\authors{R.~L.~Bailey\affil{1}, R.~Leonhardt\affil{1}, C.~M\"ostl\affil{2}, C.~Beggan\affil{3}, M.~A.~Reiss\affil{2}, A.~Bhaskar\affil{4}, A.~J.~Weiss\affil{2,5}}


\affiliation{1}{Conrad observatory, Zentralanstalt f\"ur Meteorologie und Geodynamik, Vienna, Austria}
\affiliation{2}{Space Research Institute, Austrian Academy of Sciences, Graz, Austria}
\affiliation{3}{British Geological Survey, Edinburgh, UK }
\affiliation{4}{Space Physics Laboratory, ISRO/Vikram Sarabhai Space Centre, Trivandrum, India}
\affiliation{5}{Institute of Physics, University of Graz, Universit\"atsplatz 5, 8010 Graz, Austria}

\correspondingauthor{Rachel Bailey}{r.bailey@zamg.ac.at}

\begin{keypoints}
\item \change{We aim}{The aim is} to directly forecast GICs rather than $dB/dt$, which is often used as a proxy.
\item \remove{We compare} Results from LSTMs predicting either Ex and Ey or substation GICs from solar wind data \add{are compared}.
\item GIC forecasting seems to work best when the LSTM model is trained directly on GIC data.
\end{keypoints}



\begin{abstract}
The forecasting of local GIC effects has largely relied on the forecasting of $dB/dt$ as a proxy and, to date, little attention has been paid to directly forecasting the geoelectric field or GICs themselves. We approach this problem with machine learning tools, specifically recurrent neural networks or LSTMs by taking solar wind observations as input and training the models to predict two different kinds of output: first, the geoelectric field components \Ex{} and \Ey{}; and second, the GICs in specific substations in Austria. The training is carried out on the geoelectric field and GICs modelled from 26 years of one-minute geomagnetic field measurements, and results are compared to GIC measurements from recent years. The GICs are generally predicted better by an LSTM trained on values from a specific substation, but only a fraction of the largest GICs are correctly predicted. This model had a correlation with measurements of around 0.6, and a root-mean-square error of 0.7~A. The probability of detecting mild activity in GICs is around 50\%, and 15\% for larger GICs.
\end{abstract}

\section*{Plain Language Summary}
Using satellites, we measure the state of the solar wind a short distance away from the Earth (at the so-called Lagrange-1 or L1 point) to see what is coming towards us at any given moment. Changes in the solar wind such as an increase in wind speed or a strong magnetic field can potentially impact satellite operation in orbit and power grid infrastructure on the ground - in extreme cases, solar storms can damage power grids and transformers by inducing electrical currents in the power lines. These are called geomagnetically induced currents (GICs). Here, we attempt to forecast the scales of GICs by applying machine learning methods, specifically Long-Short-Term-Memory recurrent neural networks, to take the solar wind data measured at the L1 point and predict the currents that would be seen in power grids in Austria. This gives us a lead time of around 10 to 40 minutes in the forecast. We discuss whether it is best to attempt to predict the ground electric field that leads to the GICs or the GICs themselves, and discuss the difficulties in this kind of prediction and the shortfalls in the model.

\section{Introduction}




Geomagnetically induced currents (GICs) have long been known to affect power grids, transformers and any earthed conductive networks spanning large distances \cite<for an overview, see>{Boteler1998, Boteler2017, Kelbert2020}. GICs can cause problems in power grid operation such as transformer overheating or permanent transformer damage and system collapse in extreme cases \cite{Molinski2002}, leading to further societal and economic harm \cite{Eastwood2018}.
Although studies of GICs were restricted to high latitudes where the consequences are more pronounced, mid-latitudes are being paid increasingly more attention as local effects such as transformer overheating are discovered \cite[among others]{Barbosa2015, Butala2017, Lotz2017, Gil2019, Caraballo2020, Svanda2020}.

The forecasting of GICs has developed alongside studies into the effects of regional GICs \cite{Pulkkinen2006}. Forecasting in particular is a complex problem due to the chain of cascading induction effects from the impingement of solar wind at the bow shock down to currents flowing between the earth and power grids on the surface. Improving predictive GIC modelling is listed as one of the open questions still to address to achieve GIC readiness \cite{Pulkkinen2017}. 

Most studies so far have focused on predicting geomagnetic activity - such as $dB/dt$, which is often used as a proxy for GICs - from solar wind data measured at L1 or in near-Earth space. The earliest studies addressing this problem with neural network architecture are \citeA{Wintoft2005} and \citeA{Wintoft2015}, followed by \citeA{Lotz2015} and recently \citeA{Keesee2020} and \citeA{TasistroHart2020}. The $Dst$/$SYMH$ index in particular has received a lot of attention from geophysicists and machine learning engineers alike \cite<e.g.>{Lu2016, Bhaskar2019, Wintoft2021}.

While $dB/dt$ is often used as a proxy for GICs, it does not provide the whole picture. The downside of modelling with this approach is that $dB/dt$ only functions as a useful indicator of GIC activity.
The relationship between $dB/dt$ and E (which is the primary factor determining the scale of the GICs) depends on the magnetotelluric transfer function, which is frequency dependent \cite{Chave2012MTBook}.
Single values of the time derivative of the magnetic field can only be useful GIC proxies if further assumptions on the frequency content are made \cite{Pulkkinen2006}.

What do we do if we want to develop a model that provides forecasts that power grid operators can work with? One approach would be to directly forecast the surface geoelectric field, from which GICs at different stations can be calculated. In comparison to the many studies into forecasting $dB/dt$ and $Dst$, little effort has been devoted to forecasting geoelectric fields thus far. \citeA{Pulkkinen2009, Pulkkinen2010} studied the forecasting of GICs from remote solar observations, allowing a few days warning before larger events. Modelling of geoelectric fields from solar wind to ground using full MHD modelling has been carried out by \citeA{Pulkkinen2007a}, \citeA{Zhang2015} and \citeA{Honkonen2018}, and with empirical modelling in \citeA{Lotz2017a}. 

In this study, we aim to tackle this problem from another angle and forecast regional GICs from L1 solar wind data using a machine learning method, and we compare the results to observations of GICs in Austria. We try this with two different approaches: in the first, we train a model to forecast the geoelectric field and calculate the GICs from there, and in the second we forecast the GICs directly. Predictions from both methods are evaluated and compared using data from recent years.

This study is structured as follows. Section \ref{sec:data} describes the data used in this study, including an analysis of 26 years of geomagnetic measurements used to model GICs in the region of Austria and a case study looking at the 2003 Halloween storm. Section \ref{sec:ml} then goes on to describe the models built to forecast GIC values, and the results are presented in Section \ref{sec:results}, discussed in Section \ref{sec:discussion} and summarised in Section \ref{sec:summary}.

\section{Data} \label{sec:data}

This analysis relies on INTERMAGNET-quality geomagnetic observatory data, which ensures a high quality of data with few data gaps or disturbances. We use data with a cadence of one minute because these are available for a long time period (26 years), which is not possible with 1 Hz data. Data with 1-minute resolution should be representative of most important GIC content \cite{Pulkkinen2006}. Due to Austria's small size (roughly 280 x 600 km), we assume that the geomagnetic variations are roughly constant across it both latitudinally and longitudinally, and therefore only select and use geomagnetic variations from one station at a time. 

In the following, we describe the data sets used in this study. Geomagnetic field variations from observatory measurements were used to calculate the ground geoelectric field in Austria. GICs at any power grid substation can be calculated from the geoelectric field, and the equations for two specific substations are determined using a linear fit to observed GICs. In terms of the geomagnetic and geoelectric field components, $x$ and $y$ refer to the geographic northward and eastward directions respectively.

\subsection{Geomagnetic observatory data from WIC and FUR}


The Conrad Observatory (WIC), situated at a geomagnetic latitude of 42.95$^{\circ}$ and longitude of 89.94$^{\circ}$ according to AACGM-v2 \cite{Shepherd2014}, is located southwest of Vienna near the town of Muggendorf in Lower Austria. High quality geomagnetic measurements have been carried out here since the official opening mid-2014, providing six years of data for analysis. We extend the time range using data from F\"urstenfeldbruck (FUR) in Bavaria, Germany. Initial studies are done using WIC data, and studies of long-term measurements are carried out using FUR data. A map showing the location of the two stations can be found in \textbf{Fig. \ref{fig:map}}.

The F\"urstenfeldbruck Geomagnetic Observatory (geomagnetic lat: 43.06$^{\circ}$, lon: 85.93$^{\circ}$) is one of the closest INTERMAGNET-quality geomagnetic observatories to the Conrad Observatory. It is situated almost directly west of WIC and separated by $348$ km. This station is a very good proxy for geomagnetic field variations in Austria due to its proximity and the similar geomagnetic latitude and geological setting. Measurements at a quality high enough for this analysis have been carried out since 1995, providing twenty-six years of data or 13.7 million data points at a 1-minute resolution.

An analysis of the coherence between WIC and FUR data has been carried out for the overlapping years of measurements (2015-2021), in which the Pearson's correlation coefficient (PCC) between the two time series doesn't drop below $0.99$ for either the $x$ or $y$ variables over all six years. The correlation in variations ($dBx/dt$ and $dBy/dt$) is slightly lower, with the lowest values ($0.91$) seen in the $dBy/dt$ values.

\subsection{Geoelectric field} \label{sec:e-field}

In order to model the expected levels of GICs, we need knowledge of the ground geoelectric field in the region. The geoelectric field for the past 26 years is modelled directly from the 1-minute geomagnetic field variations at FUR. The model approach used is the one-dimensional plane wave method \cite<e.g.>{Boteler2017} using the EURHOM model number 39 \cite{Adam2012} to describe the one-dimensional layers of resistivity going into the Earth. We assume the time series is representative across the country, which is a reasonable approach for small areas but not for larger countries. The plane wave approach was used in favour of the thin-sheet approach used in previous studies \cite{Bailey2017, Bailey2018} for the shorter computation times with similar levels of accuracy. The calculation results in the horizontal geoelectric field components \Ex{} and \Ey{}. Note that the $x$-component in the geoelectric field corresponds to the $y$-component geomagnetic field variations, and vice versa.

\subsection{Geomagnetically induced currents} \label{sec:gics}

\begin{figure}[t]
    \centering
    \includegraphics[width=\textwidth]{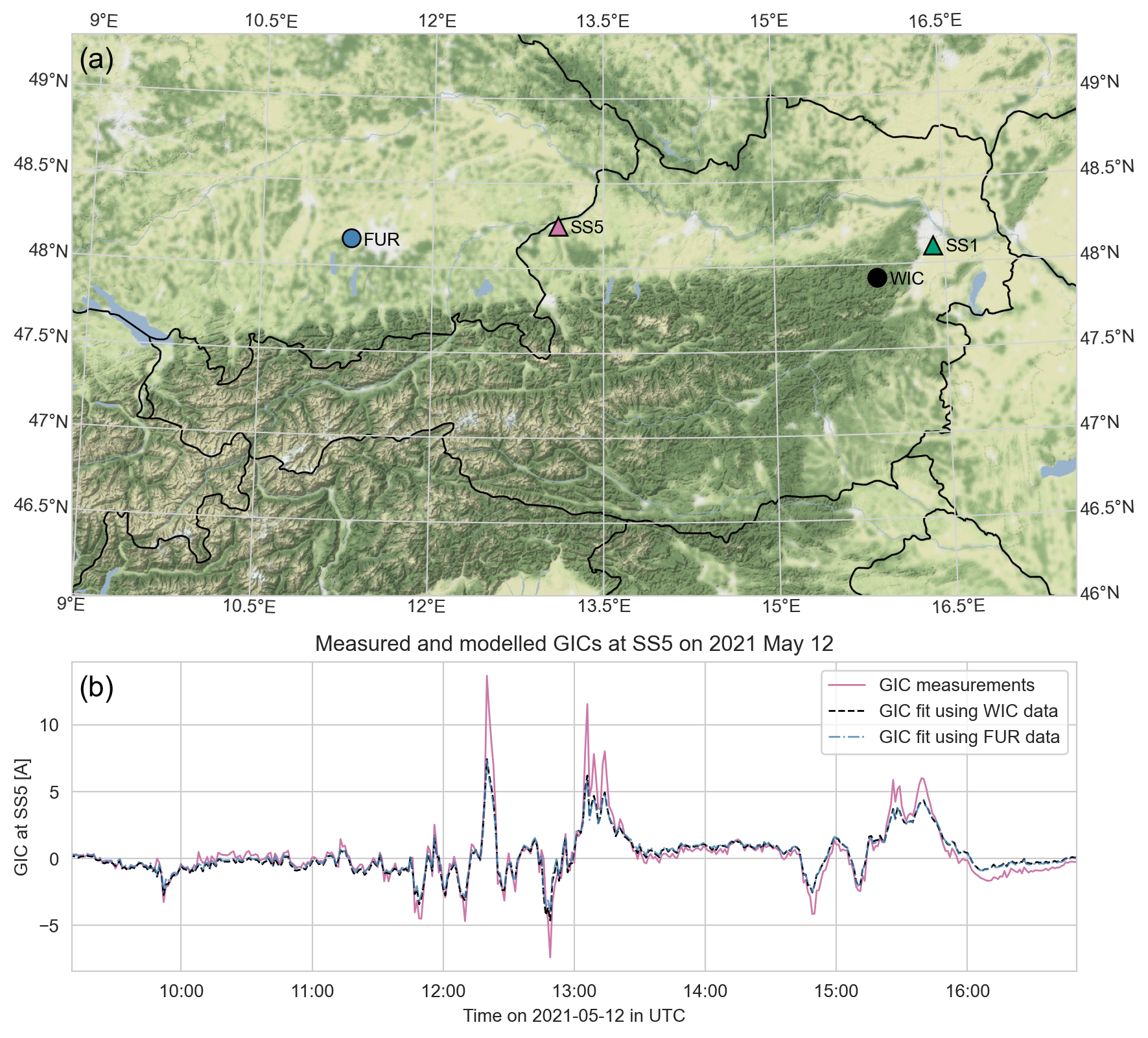}
    \caption{(a) A map showing the locations of two power grid substations (triangles) and the two geophysical observatories (circles) used for geoelectric field modelling, and (b) an example of GIC fit from modelled geoelectric field values for a geomagnetic storm in May 2021. The solid line (purple) shows transformer neutral point current measurements that have been offset-corrected and resampled via interpolation to a 1-minute sampling rate (from 1-second). The two dashed lines show the GICs calculated from E using WIC (black) and FUR (blue) data, which are nearly identical. Note that the largest GIC values are almost always underestimated despite the otherwise good agreement between model and measurements.}
    \label{fig:map}
\end{figure}

To evaluate the levels of GICs over the 26 years of available FUR data, we do not follow the standard modelling procedure of putting the geoelectric field components through the full power grid network, which would be computationally heavy, but instead find a direct linear fit of the geoelectric field components to measurements of GICs to find the current at station $j$, i.e.

\begin{equation} \label{eq:gicfit}
    GIC_{j} = a_j \cdot E_x + b_j \cdot E_y
\end{equation}

\noindent where $a_j$ and $b_j$ are station-specific real coefficients (with units A$\cdot$km/V). This approach can only be used on transformer stations with measurements\add{ since the coefficients must be determined from a linear fit to the data}, but it often has similar or better accuracy than results from a network model. See \citeA{Pulkkinen2007} or \citeA{Torta2012} for more discussion on this method and for the equations determining $a_j$ and $b_j$.

The fit for Eq.~\ref{eq:gicfit} was applied to measurements of direct currents from multiple transformer neutral points in Austrian power grid substations provided by the Graz University of Technology\add{, a summary of which can be found in} \citeA{Albert2022}. In this study, only measurements from two substations were used: one near Vienna (hereafter referred to SS1 for Substation 1) and another north of Salzburg (SS5)\add{, both with sampling rates of one second. The data was resampled to a one minute sampling rate for use in this study using a 1-minute median sliding window}. These two stations are of interest because they are in the high-voltage network and experience larger GICs than the other stations with measurements. As such they are useful examples for depicting the expected maximum scales of GICs that could be seen across the grid. We choose three geomagnetically active periods and use the geoelectric field components \Ex{} and \Ey{} modelled from FUR data to derive the following equations:

\begin{equation} \label{ss1-eq}
    GIC_{SS1} = 3.77 \cdot 10^{-2} \cdot E_x + 3.19 \cdot 10^{-2} \cdot E_y
\end{equation}

\begin{equation} \label{SS5-eq}
    GIC_{SS5} = 0.44 \cdot 10^{-2} \cdot E_x + 5.55 \cdot 10^{-2} \cdot E_y
\end{equation}

We see that the $x$-component of the geoelectric field contributes roughly the same amount to the GICs seen in SS1 as the $y$ component. The $y$-component of the geoelectric field mostly dominates the currents in SS5 and contributes ten times more than the $x$-component. The differences in contributions from geoelectric field components stem from the varying grid layout and connections at each substation. An analysis shows that the GICs calculated from these equations are slightly more accurate than those from the full network model. Comparing to measurements at SS1, the Pearson's correlation coefficients for both GICs from the network model and GICs from Eq.~\ref{eq:gicfit} are $0.86$, while at SS5 the correlation improves from $0.85$ to $0.88$. In both cases the amplitudes of the GICs are better matched and the root-mean-square-errors drop from $0.24$ to $0.12$~A at SS1 and $0.46$ to $0.12$~A at SS5. These measures \change{are for eight days}{were calculated from a fit of the geoelectric field data to measurements using eight days} of geomagnetically active periods (including the September 2017 storm)\remove{ used to find the coefficients}. \add{This includes the most recent active period, meaning the measurements should represent the current grid configuration and we exclude fitting only to grid noise by using a geomagnetically active period.} A fit applied to the geoelectric field modelled from WIC rather than FUR data produces slightly different coefficients but results in the same level of accuracy when compared to GIC measurements. An example of the measurements and GIC fits can be seen in \textbf{Fig. \ref{fig:map}b}.

Regardless of which time range the fit is applied to, the GICs calculated using Eq. \ref{eq:gicfit} (as well as those from the network model) tend to underestimate the peaks of the largest GICs by up to a factor of two (see e.g. \textbf{Fig \ref{fig:map}b}, 12:20 or 13:05 UTC). We assume this is a result of attenuation of the modelled geoelectric field due to the lower sampling rate used for field modelling \cite{Grawe2018} or the oversimplification of using a uniform geoelectric field and 1D model of the subsurface resistivity \cite{Ngwira2015, Sun2019, Weigel2017}. Despite this, the very good agreement between model and measurements means that any results based on the modelled geoelectric fields will still be reasonable.

In addition to the absolute GIC values, we also look at the cumulative absolute GICs over an hour, \GICsum{}. \GICsum{} is taken as the sum of values over the hour\add[]{ divided by the number of timesteps in an hour (60 for our minute values) to make it independent of sampling rate, and is used} as a separate indicator for geomagnetic activity, more representative of sustained GICs than large spikes, both of which can have different (but similarly detrimental) effects on transformers \cite{Bolduc2002, Gaunt2007}. Using the accumulated sum of GICs or geoelectric field has seen usage in other studies, although not often - \citeA{Lotz2017} used the accumulated E over varying periods and \citeA{Viljanen2014} also worked with daily GIC sum averaged across nodes. \change{The scale of GICsum will vary depending on the sampling rate of the data used, but in the case of minute data i}{I}n Austria, \change{0 to 50}{0 to 0.5}~Ah can be seen during quiet times, and values above that generally represent more active times.

\subsection{Distribution of values}

\begin{figure}[t]
    \centering
    \includegraphics[width=\textwidth]{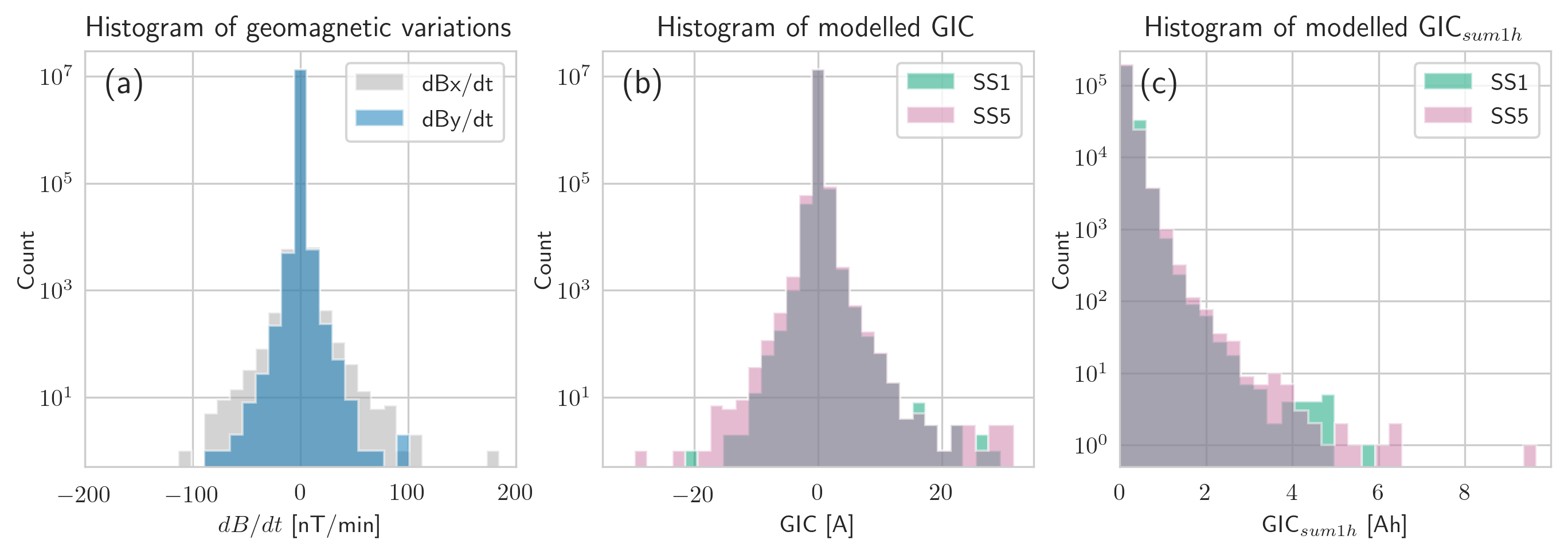}
    \caption{Histograms showing the distribution of the values in (a) the geomagnetic variations at FUR, (b) the GICs modelled from $dB/dt$ at two substations, and (c) the hourly cumulative modelled GICs at two substations for all data, \GICsum{}. The y-axes have logarithmic scales.}
    \label{fig:hist_all}
\end{figure}

In order to determine how best to forecast GICs, we first look at the 26 years of available data and the distributions of both geomagnetic variations and modelled GICs. \textbf{Figure \ref{fig:hist_all}a} presents the distribution of FUR minute $dBx/dt$ and $dBy/dt$ variations. There are very few values populating the tail of the distribution where the largest values are found. High values for this region are at $80$ nT/min and upwards. The largest variations occur most commonly in the $x$-direction (leading to larger \Ey{}) rather than the $y$-direction, implying that stations in the power grid sitting on east-west lines are already more susceptible to larger GICs.

In \textbf{Figs. \ref{fig:hist_all}b and 2c}, the GICs observed at SS5 are larger than those at SS1. While the size of the currents depends largely on the network topology and grounding resistance, we noted in Section \ref{sec:gics} that the currents at SS5 are mostly determined by the $y$-component of the geoelectric field (or $x$-component of the geomagnetic field variations), which generally sees larger variations.

\subsection{Most active days}

\begin{table}[]
\caption{Table showing the ten most active days according to the maximum values in three measures: leftmost are the horizontal geomagnetic field variations ($dBx/dt$ and $dBy/dt$), in the centre the absolute GICs (\GICabs{}) at two different transformer stations (SS1 and SS5), and rightmost the cumulative GICs over an hour at two transformer stations (\GICsum{}). Bold font highlights the ten largest values seen in that measure. The largest values are seen during the Halloween Storm on 2003 October 29-31 (italicised).} \label{tab:active-days}
\begin{tabular}{l|llllll}
Date       & $dBx/dt$       & $dBy/dt$      & \GICabs{1} & \GICabs{5} & \GICsum{1} & \GICsum{5} \\
           & [nT/min]       & [nT/min]      & [A]             & [A]             & [Ah]             & [Ah]               \\ \hline
1998-05-04 & 52.0           & \textbf{46.0} & \textbf{11.37}  & 9.56            & 2.48              & 2.81              \\
2000-04-06 & 42.9           & \textbf{43.7} & 8.78            & 11.34           & 3.00              & 3.43              \\
2000-07-15 & \textbf{184.7} & 28.5          & \textbf{20.30}  & \textbf{29.47}  & \textbf{4.79}     & \textbf{6.25}     \\
2000-09-17 & 34.5           & 19.9          & 10.45           & 9.89            & \textbf{4.21}     & \textbf{4.21}     \\
2001-03-31 & \textbf{82.4}  & \textbf{40.7} & 10.85           & \textbf{17.55}  & 3.69              & 3.18              \\
2001-11-06 & \textbf{85.1}  & \textbf{38.1} & \textbf{12.73}  & \textbf{13.72}  & 3.95              & \textbf{5.24}     \\
2001-11-24 & \textbf{62.4}  & 33.3          & \textbf{14.20}  & \textbf{17.81}  & \textbf{4.51}     & \textbf{4.18}     \\
\textit{2003-10-29} & \textbf{102.9} & \textbf{92.3} & \textbf{28.57}  & \textbf{31.67}  & \textbf{5.77}     & \textbf{9.66}     \\
\textit{2003-10-30} & 33.1           & \textbf{40.3} & \textbf{17.68}  & \textbf{16.44}  & \textbf{4.66}     & \textbf{4.71}     \\
\textit{2003-10-31} & \textbf{91.5}  & \textbf{56.2} & \textbf{14.75}  & \textbf{16.88}  & 2.41              & \textbf{4.03}     \\
2003-11-20 & 19.8           & 31.4          & \textbf{11.63}  & 10.73           & \textbf{4.82}     & \textbf{4.62}     \\
2004-07-26 & \textbf{78.5}  & 8.5           & 10.15           & \textbf{15.33}  & 1.20              & 1.53              \\
2004-11-07 & 43.0           & \textbf{37.7} & 7.33            & 8.60            & 2.54              & 2.67              \\
2004-11-08 & 24.7           & 28.9          & 9.77            & 9.42            & \textbf{4.17}     & 3.57              \\
2004-11-09 & \textbf{76.1}  & \textbf{49.9} & \textbf{14.21}  & 13.70           & \textbf{4.28}     & 3.46              \\
2005-05-15 & 36.3           & \textbf{35.1} & \textbf{11.45}  & \textbf{13.96}  & \textbf{4.38}     & \textbf{6.31}     \\
2005-08-24 & 41.6           & 31.9          & 10.51           & 13.18           & \textbf{4.01}     & \textbf{6.16}     \\
2005-09-11 & \textbf{60.7}  & 30.7          & 8.81            & 12.50           & 1.24              & 1.65              \\
2015-06-22 & \textbf{63.0}  & 12.8          & 9.94            & \textbf{16.67}  & 2.56              & 3.47             
\end{tabular}
\end{table}

In \textbf{Table \ref{tab:active-days}}, the 10 most active days in the 26 years of data according to different measures of activity $dBx/dt$ and $dBy/dt$ at FUR, modelled \GICabs{} and \GICsum{} at both SS1 and SS5 are listed. There are many overlapping days between the different measures, making a total of 19 days. Bold font highlights the ten largest values in each column.

A similar table for largest GIC days in Central Europe was produced in \citeA<>[Table 4]{Viljanen2014}, and we see that the tables are very much in agreement with 17 shared dates, even though the table in \citeA{Viljanen2014} is only based on one variable. They used a value akin to the \GICsum{} used here, namely the daily sum of GICs averaged across all nodes. Similarly, 17 of the days listed here also appear in \citeA{Juusola2015}, Table 3, where an analysis of the days with largest GICs was carried out for Northern Europe. Other larger storms that have occurred since those studies (March 2015 and September 2017) do not stand out in comparison to those from the last solar cycle with the exception of the storm from June 2015.

\begin{figure}
    \centering
    \includegraphics[width=\textwidth]{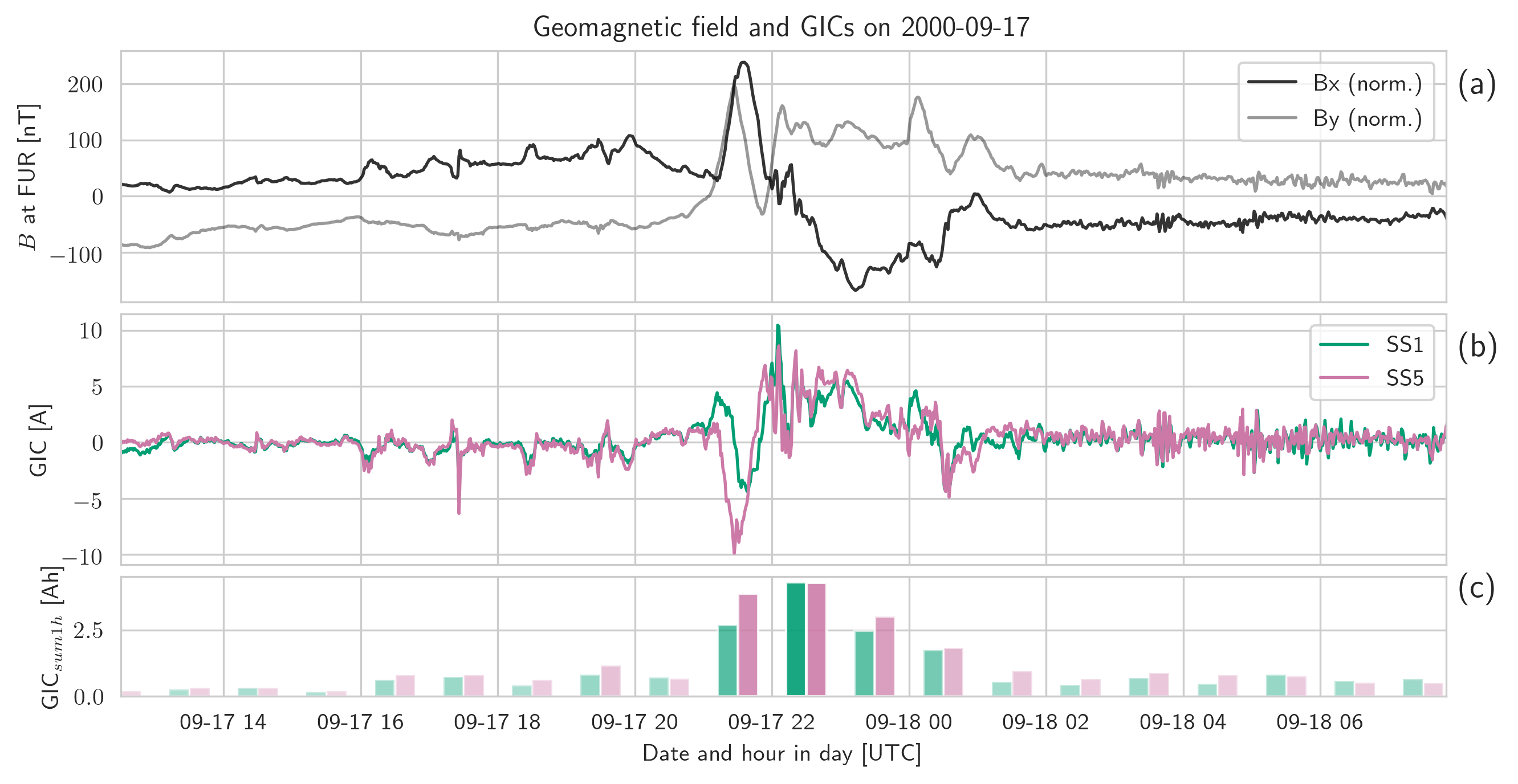}
    \caption{Plot of (a) geomagnetic variations at FUR (normalised to around zero by subtracting the mean field strength), (b) modelled GICs at two substations, and (c) cumulative hourly GICs on 2000-09-17 as an example of a day with no extreme GIC values but large cumulative hourly GICs.}
    \label{fig:gicsum_example}
\end{figure}

The largest values in each measure are clearly centered around the 2003 Halloween storm. Large values in $dBx/dt$ tend to go alongside large GIC values in SS5, and days with large \GICsum{} usually coincide with days with larger \GICabs{}, as expected. Some exceptions are 2000-09-17, 2001-04-08, 2005-01-07 and 2005-08-24, which only show high cumulative GICs but do not stand out in $dB/dt$-values and peak GICs. A comparison of these events shows they have large and unidirectional geomagnetic field variations (with total field changes of $100$ to $300$ nT) that occur over an hour or more. These in particular lead to sustained GICs in stations susceptible to geomagnetic field changes in that direction. The variations on 2000-09-17 are shown as an example of this kind of behaviour in \textbf{Fig. \ref{fig:gicsum_example}}. Although not extremely geomagnetically active, they show that power grid transformers would have been subjected to large amounts of cumulative GICs sustained over an hour at least.

\subsection{Case study: 2003 Halloween Storm} \label{sec:ha}

In \textbf{Fig. \ref{fig:hist_all}}, almost all of the values in the tail end of the distribution resulted from the ``Halloween storm", which lasted from 2003 October 29 to November 1. These also make up the largest values in \textbf{Table \ref{tab:active-days}}, with maximum GIC values almost twice as large as the other values seen. We now conduct a detailed analysis of the behaviour during this storm and the GICs that were likely present in the power grid as an example of the problems that can arise when using only $dB/dt$ as a proxy for GICs. We see that both large instantaneous GICs and sustained GICs appear without large $dB/dt$ values.

The geomagnetic storm that occurred at the end of October in 2003 was the result of a series of fast and geoeffective coronal mass ejections hitting the Earth during a particularly active period around the maximum of solar cycle 23 \cite<e.g.,>{gopalswamy2005}. In \citeA{Eastwood2018}, this storm was classified as a 1-in-10 year event, and is not considered an exceptionally rare example. No event of this or a higher magnitude has occurred since 2003 \cite<with the exception of a CME directed away from Earth on July 2012, see>{Ngwira2013,baker2013,liu2014}, and such events are somewhat more probable during the solar maxima \cite{owens2021}, but have also occurred at any point throughout the solar cycle.  

\begin{figure}
    \centering
    \includegraphics[width=0.95\textwidth]{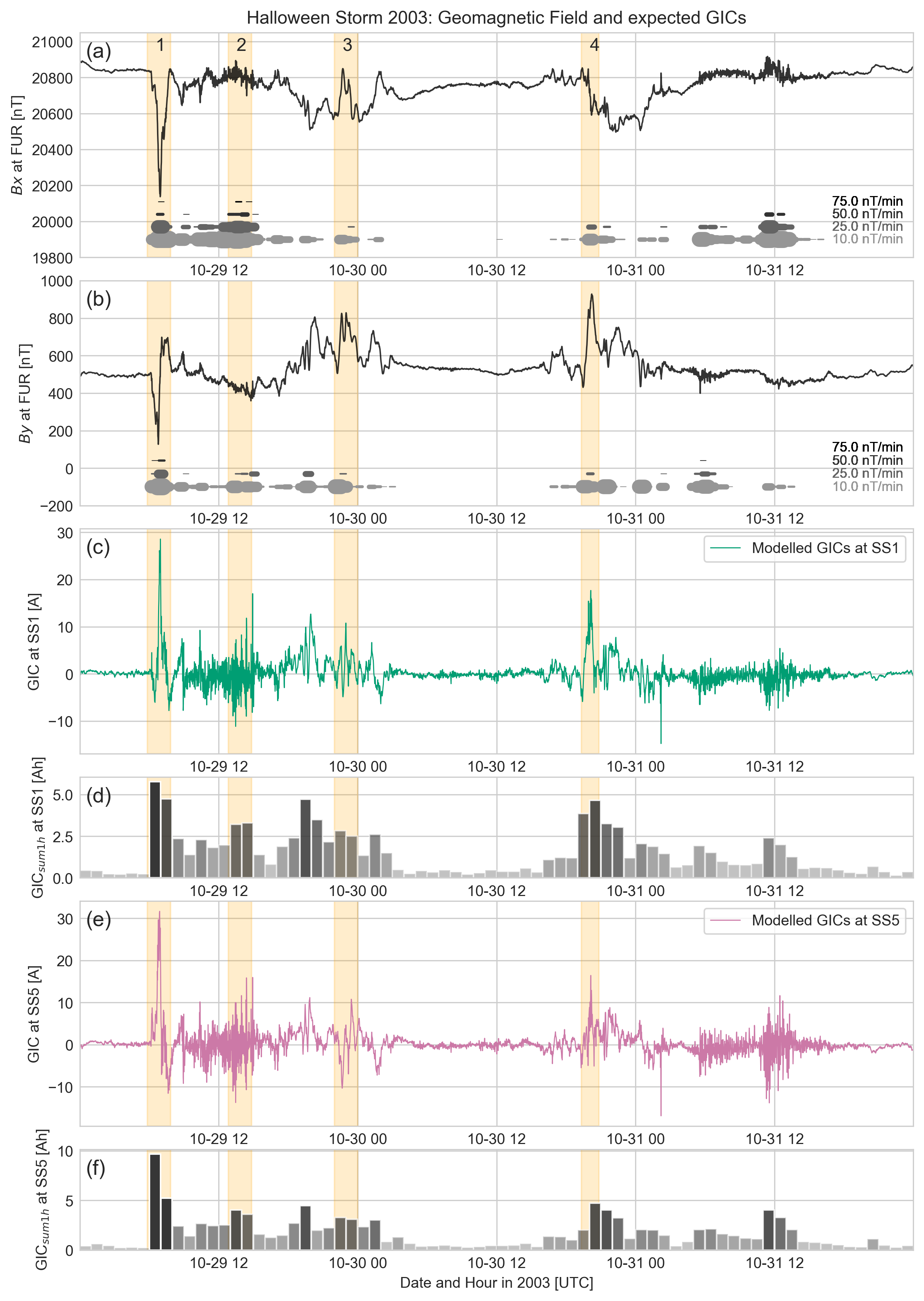}
    \caption{The Halloween storm from 2003 October 29 till 2003 November 1, during which some of the largest geomagnetic variations of the last few decades were seen. (a) and (b) show the geomagnetic variations at FUR in the $x$ and $y$ directions. Plotted below are levels of activity (10, 25, 50, and 75 nT/min) with line thickness showing how often these values were exceeded over a certain time range. (c) and (e) show the modelled GICs at the substations SS1 and SS5, and (d) and (f) show the cumulative GICs over each hour at each substation.}
    \label{fig:halloween}
\end{figure}

A brief evaluation of this storm for Austria was carried out in \citeA{Bailey2018}, in which a maximum GIC of $14$ A was modelled. Using an updated model with newer data allows us to get a more accurate estimate of GICs during stronger events, and using the method from Section \ref{sec:gics} for SS1 and SS5 we see the values reaching $25-30$ A. Taking into account that the GIC peaks modelled using minute data generally underestimate the observations, these could also have reached up to $60$ A.

\textbf{Figure \ref{fig:halloween}} compares the geomagnetic field and the modelled GICs for the 2003 Halloween Storm. Panels (a) and (b) show the geomagnetic field variations in the $x$ and $y$ directions. The thick lines plotted below the field show the presence of various levels of $dB/dt$ variations (as they might be shown using a forecasting method). Light grey shows a level of $10$ nT/min, and this increases going upwards to $25$ nT/min, $50$ nT/min and $75$ nT/min. The thickness of the line shows how often the value was exceeded within a time frame of 30 minutes (with a maximum being 30 times). Panels (c) and (e) show the GICs calculated from the modelled geoelectric field at the substations SS1 and SS5, and the panels (d) and (f) show the cumulative sum of absolute GIC values (\GICsum{}) over 1-hour periods.

Four time intervals, highlighted in yellow on the plot, have been picked out for discussion. Intervals 1 and 2 have been selected because, as can be seen in the high levels of $dB/dt$ in both components, these were the most active periods. Intervals 3 and 4, in contrast, were chosen because of continuously low levels of $dB/dt$ but lack of higher ($>50$ nT/min) values. 

Interval 1 shows a large GIC value, which is fairly short-lived. Interval 2, in contrast, shows a consistent level of moderate GICs, though it does not reach an extremely high value. Interval 3 has a similar level of sustained \GICsum{} as Interval 2 despite it having a comparatively smaller amount of $dB/dt$ over the same period. In Interval 4, SS1 experiences the second highest value of GIC ($17$ A) throughout the whole storm, even though there is only continuous low-level $dBx/dt$ and $dBy/dt$ (10 to 25~nT/min), most of it unidirectional (comparable to the type of signal seen in \textbf{Fig. \ref{fig:gicsum_example}}). On top of that, the cumulative GICs are also some of the highest.

In summary, we see there are large differences between periods that have short-lived but large GICs (Intervals 1 and 4) and those that have longer periods of sustained GICs (Intervals 2 and 3), and both large GICs and sustained GICs can appear without large $dB/dt$ because the ground geoelectric field responds at a range of frequencies not captured by $dB/dt$ intensity alone. Each scenario could lead to different problems if it were to occur in a transformer to any large degree \cite{Price2002, Gaunt2007, Bolduc2002}.

\section{Building a Forecasting Model} \label{sec:ml}

From the analysis of past data, we deduce that, in order to forecast a comprehensive summary of expected GIC behaviour, we need to forecast either both geoelectric field components or the GICs directly. While the magnitude of the field is most important, the direction also plays an important role. From Eqs. \ref{ss1-eq} and \ref{SS5-eq}, we see that a large value in \Ex{} at SS5, for example, could be cancelled out by a smaller negative one in the \Ey{} value, and the opposite could be true elsewhere, making a station-by-station approach advantageous.

We now move on to build a forecasting model based on these conclusions. Three machine learning methods were put through an initial comparison for evaluation: a standard feed-forward neural network (NN) with three layers (32 neurons initially), a gradient boosting regressor based on XGBoost in Python (with 400 decision trees), and a recurrent neural network (specifically, a Long-Short-Term Memory RNN or LSTM) with three layers (32 blocks initially) and a basic Attention mechanism. \change[]{The models were}{The three types or architecture were set up in size and hyperparameter choice to be somewhat comparable in basic accuracy on an initial subset of the training data set, then were provided the full, identical data sets (scaled and shaped according to each method) and} compared according to a set of metrics for model evaluation (root-mean-square error, Pearson's correlation coefficient, probability of detection). From these first comparisons, the LSTM with Attention showed the most promise and was developed into the final model, although due to the myriad machine learning methods available these days there may well be other approaches equally suited for this task. \add[]{Details on the comparison can be found in the Jupyter Notebook \#4 listed in Sec.} \ref{sec:datasources}.

\subsection{Data preparation}

The input to the machine learning model is solar wind data measured at L1 and forward-propagated to the bow shock. This means\add{ that, assuming we take measurements from satellites situated at L1,} we have a varying forecast lead time between 15 and 60 minutes depending on the solar wind speed. The high resolution OMNI data set (see section on Data Availability for details) was used for solar wind measurements (speed, density, and magnetic field components) at a minute cadence\add{ combined with the local time and day in year} to make up the features, while the model target was either the geoelectric field (E) modelled from FUR data or the GICs modelled from the \Ex{} and \Ey{} components.

Taking solar wind measurements that have already been propagated forward to the bow shock, we use the two hours prior to the time we wish to forecast as input. This goes from $t-120$ minutes to $t-0$, where $t$ is the forecast time. The range of 120 minutes for past data was decided on through experimentation, where the period was increased until longer periods did not lead to any improvements in the forecasting skill. To \change{reduce noise in}{reduce the size and complexity of} the input data, it is subsampled to a 10-minute resolution by picking every 10th point (\change{we found}{rather than} interpolation and/or fitting\add{, which we found} led to a loss in forecast skill), resulting in sequences of length 12. These sequences are used as input to forecast the maximum value of E or GICs over 40 minutes from $t-10$ to $t+30$. This step of ten minutes into the ``past'' (which reduces the lead time by ten minutes) is to account for possible timing errors in propagating the solar wind forward to the bow shock.

Sampling the modelled geoelectric field or GIC data to produce a balanced data set for model training is challenging because there is a clear bias towards quiet times and not enough data from geomagnetically active times (with a factor of roughly $10^7 : 1$ for quiet to active). \add{An initial approach using the entire data set led to a trained model that predicted only quiet times, which could not be remedied without additional data handling or large changes to the training methods.} The target data set was therefore selectively sampled to reduce the imbalance. The distribution of samples was undersampled in the range of E $= 0$ to $100$ mV/km (GIC $= 0$ to $8$ A). Above that, we applied some data augmentation by duplicating the samples by 2 to 5 times and applying a random offset in time to the input data of each to avoid identical samples. \add{The offset was randomly sampled without replacement from values between -10 and +10 minutes, which shifts the input solar wind data that the model sees, and means that the maximum value was either closer to the start or the end of the following 40-minute forecast window. }Otherwise, all samples had a minimum time difference of 60 minutes between them. The resulting distribution is close to a one-sided Gaussian distribution. Roughly the same number of samples (9000) were used in training for each target.

The samples were split into training and testing sets by time. The years 2000, 2001 were reserved for validation to aid in model selection during training, while 2017, 2019 and 2020 were reserved for testing, and the remaining 21 years were used in training. The presence of data gaps longer than 15 consecutive minutes in the OMNI data set led to samples being excluded from the analysis - this led to 8 to 15\% sample exclusion, depending on the years used. Data gaps shorter than 15 minutes were linearly interpolated over.

We reduced all values of E $> 200$ mV/km (GIC $> 15$ A) to 200 mV/km ($15$~A) because the larger values were only present in roughly 100 of the 13.7 million time-steps (or five to seven events in the 25-year period) and heavily skewed the distribution, in which all values were scaled between 0 and 1. Rescaling points above this limit greatly improved the level to which the model could learn the problem but also means that the maximum forecast the model can realistically produce is for $200$ mV/km.\add[]{ This was tested by evaluating a model trained on data clipped at 200 mV/km versus one trained on the original data, and the model trained on clipped data performed better on both clipped and unclipped test data sets.}




\subsection{Training the LSTM}

To approach this forecasting problem, we use a four-layer LSTM with an Attention layer. The Attention mechanism is meant to simulate human attention \cite<first developed in>{Bahdanau2015}, which can be understood intuitively as a mechanism that picks out the most important part of a sequence and discards the parts that are considered irrelevant. It is a tool now commonly applied in natural language processing for example \cite{Galassi2020}.
The model is structured so that the input first goes through an LSTM layer and then through the Attention mechanism. The data is then fed into another LSTM layer before going through a final feed-forward layer to reduce the output to a single value.

For geoelectric field prediction, the LSTM branches into two: the left side deals with a regression problem, namely forecasting the maximum magnitude of the geoelectric field. \add{We chose a custom loss function for the regression problem where events (peaks) are rare in the data, and where the scale of the peaks is important. A min-max scaling factor used as a penalty term meant that training to match the peak value would drive the loss down.} The right side of the LSTM forecasts the sign of the geoelectric field in a classification problem\add[]{, which in this case is the sign of the maximum field value used for the regression problem}. \add{Here, the binary cross-entropy loss function was used.} Training worked better when the two were trained as separate targets, rather than attempting to forecast $E$ without taking the absolute value first. The regression problem appears to be not too difficult a task, but the model had far more problems trying to forecast the direction. In training, the weights of the two problems are, when scaled, about 15 : 1 for regression to classification. The classification problem to determine the sign is given secondary importance because even an LSTM dedicated to this problem had trouble achieving a good level of accuracy. A diagram of the different LSTM architectures\add{, the loss functions} and the hyperparameters used for the training of each model can be found in the supporting information.\add{ Iteration through the various possible hyperparameters was carried out for all four models for optimisation. Similar sets of hyperparameters were found for each LSTM application, with some minor differences between them, although the choice of the same hyperparameters for all applications also led to reasonable models in all cases. Regularisation was applied in the form of dropout.}

Multiple models were trained to evaluate the best approach for forecasting GICs. Those trained to forecast the geoelectric field components are referred to as LSTM-E, while nets trained to forecast the GICs directly are referred to as LSTM-GIC. Both neural nets are only trained on the output of geophysical models (in the case of E, the result of FUR variations put through the plane-wave model, and for GICs, these are the currents calculated in power grid transformers from E) because we don't have measurements of E or GIC over long enough periods and because, as described in Sec. \ref{sec:gics}, GICs from geophysical models reach a good enough accuracy to be a reasonable substitute in training. Both models predict the absolute value of the target, but the LSTM-E predicts the sign (positive or negative) in addition.

\subsection{Evaluating the model skill}

Each model was trained on its respective training set and the best LSTM parameters were chosen based on model behaviour when presented with the validation set. Following training, we ran the model on the test data set in a virtual `real-time mode' providing updates to the input data every 15 minutes, and giving an output with a 15-minute cadence. The comparison to the ground truth (either the modelled geoelectric field or measured GICs) is performed point-to-point as well as by looking at events, where the event-based analysis is given the most importance. In order to have a benchmark for comparison, we produced a real-time persistence approach which takes the maximum of the geoelectric field or GICs in the 20 minutes before the solar wind measurement time to forecast the maximum when the solar wind would reach Earth. As such, the persistence model (PERS) also uses a varying forecast lead time. The machine-learning forecast model should be able to beat persistence in most measures.

Our event-based analysis follows the recommendations put forward by \citeA{Pulkkinen2013} and \citeA{Welling2018} for $dB/dt$ forecasting. An ``event'' in the data is classified as a value that exceeds a certain threshold, while all values below that threshold are non-events. By defining a threshold, we can calculate the confusion matrix \cite{Wilks11}, which includes the number of correctly-predicted events or true positives (TP), missed events or false negatives (FN), incorrectly-predicted events or false positives (FP), and the correctly-predicted non-events or true negatives (TN). The metrics proposed in \citeA{Pulkkinen2013} include the Probability of Detection (POD), which is the fraction of measured events correctly predicted as events, also called the true positive rate (TPR or TP/(TP+FN)). Similarly, we include the probability of False Detection (POFD), the fraction of measured non-events incorrectly predicted as events, which is equivalent to the false positive rate (FPR or FP/(FP+TN)). In addition, the Heidke Skill Score (HSS) and True Skill Statistic (TSS) are also considered, both of which are derived from all variables in the confusion matrix \cite<see e.g.>{Heidke1926, Bloomfield2012}. Both the HSS and TSS show no model skill at 0, and better model skill when approaching 1. The TSS has the benefit over the HSS of being unbiased by event/non-event ratios. We also include the bias (BS), which shows if the model tends to over-predict (more false positives, BS $> 1$) or under-predict (more false negatives, BS $< 1$).

\section{Results} \label{sec:results}

We present the results split in two parts: in the first part, we test our model's forecasting ability with regards to the the geoelectric field components. The results are compared to the geoelectric field modelled from geomagnetic variations at FUR (see Sec. \ref{sec:e-field}). In the second part, we test the forecasting ability for GICs. These are calculated using (1) the geoelectric field components predicted from LSTM-E to calculate the GICs at the two substations we picked for analysis, and (2) directly from LSTM-GIC for each substation. The comparison between the model results and measurements of GICs is carried out for the years 2017, 2019 and 2020.

For the evaluation of geoelectric field forecast, we compute the scores for three event thresholds: these are $30$, $60$, and $90$ mV/km in both \Ex{} and \Ey{}. In GICs, the level of $60$ mV/km corresponds to a current of roughly $4$ A through either SS1 or SS5, and we use similar thresholds of $2$, $4$ and $6$ A. It is difficult to determine the minimum level of GICs above which transformers may experience adverse effects because these are heavily dependent on transformer type and the presence of DC-handling mechanisms. We have too few measurements of GICs exceeding higher levels such as $10$ A to make an analysis at this level useful, but $4$ A is crossed often during geomagnetically active times. The results are described in the next section.

\begin{figure}[htbp]
    \centering
    \includegraphics[width=1.0\textwidth]{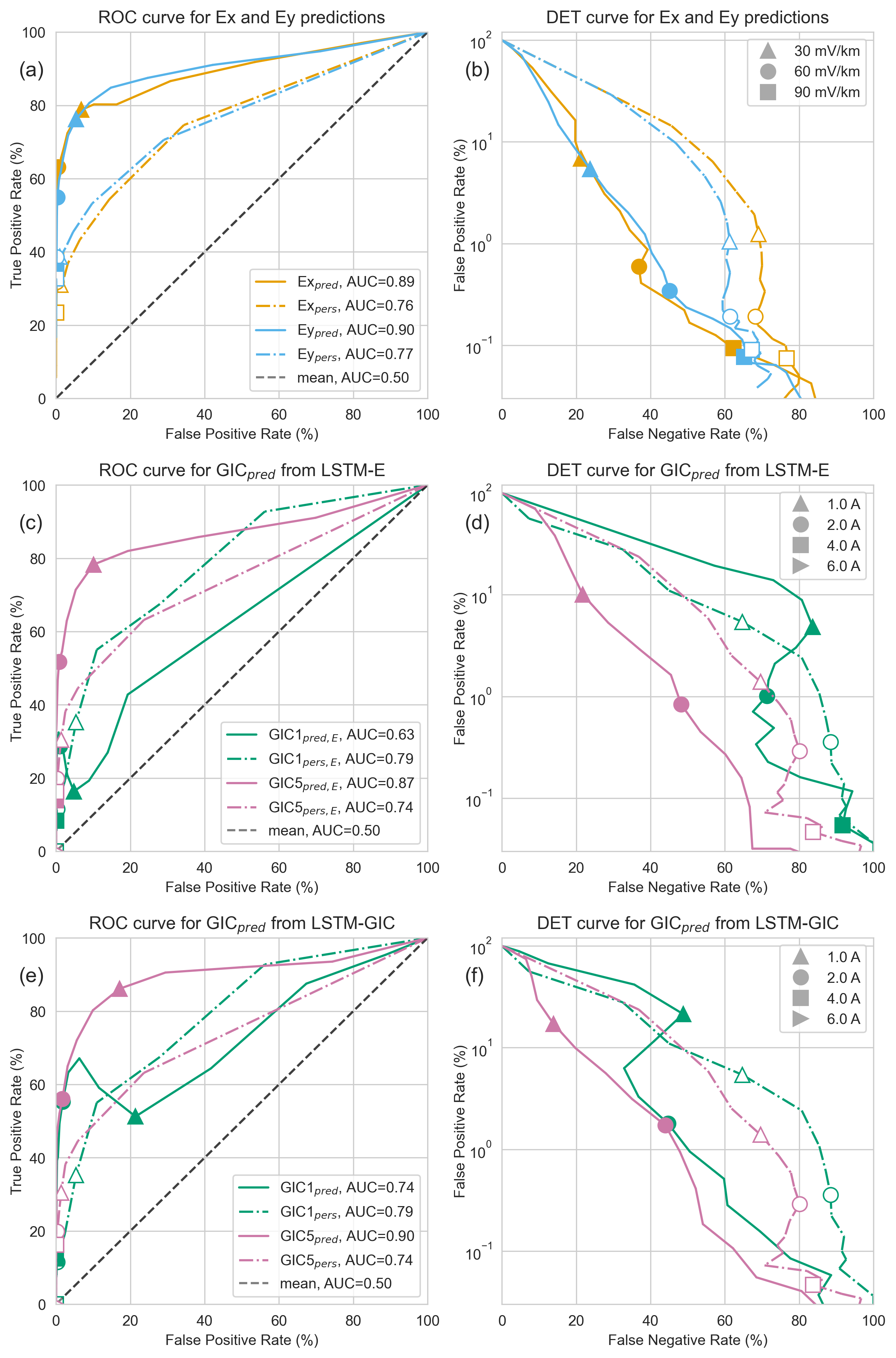}
    \caption{Receiver-operator characteristic (ROC) and detection-error tradeoff (DET) curves for three approaches: (a-b) the geoelectric field, showing the output from the LSTM-E models vs the modelled geoelectric field, (c-d) the GICs calculated from the geoelectric field predicted by LSTM-E compared to measured GICs, and (e-f) the GICs predicted by the LSTM-GIC models compared to measured GICs. SS1 and SS5 are two separate substations in the power grid from which we have measurements. The values for specific event thresholds are labelled with shapes as defined in each legend.}
    \label{fig:roc-det-all}
\end{figure}

\textbf{Figure \ref{fig:roc-det-all}} gives a graphical representation of the model behaviour at each threshold using receiver-operator characteristic (ROC) and detection-error tradeoff (DET) curves. Both depict the model's ability to forecast events at varying thresholds. The ROC curve shows the trade-off between the true positive rate (also POD) and false positive rate (also POFD) at different event thresholds. Usually, when the threshold is low, the TPR is high but we also see an increased FPR, which is unwanted - a model that captures the observed behaviour shows a curve that keeps close to the upper left corner. The area-under-the-curve (AUC in the legend) shows good model skill as it approaches 1. On the other hand, the DET curve shows the relationship between the false negative rate (fraction of all predicted non-events that were measured events misclassified as non-events, or FN/(FN+TN)) and false positive rate, the number of which usually goes up as the other goes down depending on where the threshold for an event is set. Here, the best model behaviour is seen as the curves approach the lower left corner. It is useful in error minimisation to deduce the rate at which the FNR improves with regards to an increase in FPR rate (and vice-versa).

\begin{table}[h!]
\caption{Metrics from an event-based analysis of the LSTM-E models applied to the years 2000, 2001, 2017, 2019 and 2020 in a retrospective real-time mode with the model being run at 15-minute intervals. A persistence model (PERS) is included for comparison. The first four columns provide the values for the confusion matrix (where TP, FP, TN and FN are the true positives (hits), false positives, true negatives (misses) and false negatives), the probability of detection (POD), probability of false detection (POFD), Heidke Skill Score (HSS), True Skill Score (TSS), and bias (BS). The variable TH in brackets gives the event threshold used to define events and compute the metrics.}
\begin{tabular}{l|llllllllll}
LSTM-E Model     & N$_{events,obs}$ & TP   & FP & FN   & TN     & POD  & POFD & HSS  & TSS  & BS  \\ \hline
\Epred{x}(TH=30) & 3092         & 2436 & 11749 & 656 & 160506  & 78.8 & 6.8  & 0.26 & 0.72 & 4.6 \\
\Epred{x}(TH=60) & 494          & 312  & 1038  & 182 & 173815  & 63.2 & 0.6  & 0.34 & 0.63 & 2.7 \\
\Epred{x}(TH=90) & 175          & 66   & 164   & 109 & 175008  & 37.7 & 0.1  & 0.33 & 0.38 & 1.3 \\ \hline
\Epred{y}(TH=30) & 2989         & 2279 & 9328  & 710 & 163030  & 76.2 & 5.4  & 0.29 & 0.71 & 3.9 \\
\Epred{y}(TH=60) & 559          & 307  & 600   & 252 & 174188  & 54.9 & 0.3  & 0.42 & 0.55 & 1.6 \\
\Epred{y}(TH=90) & 241          & 84   & 135   & 157 & 174971  & 34.9 & 0.1  & 0.36 & 0.35 & 0.9 \\ \hline
PERS Model       & N$_{events,obs}$ & TP   & FP   & FN & TN       & POD  & POFD & HSS  & TSS  & BS  \\ \hline
\Epers{x}(TH=30) & 3092         & 958  & 2128  & 2134 & 170127 & 31.0 & 1.2  & 0.30 & 0.30 & 1.0 \\
\Epers{x}(TH=60) & 494          & 157  & 335   & 337 & 174518  & 31.8 & 0.2  & 0.32 & 0.32 & 1.0 \\
\Epers{x}(TH=90) & 175          & 41   & 130   & 134 & 175042  & 23.4 & 0.1  & 0.24 & 0.23 & 1.0 \\ \hline
\Epers{y}(TH=30) & 2989         & 1156 & 1804  & 1833 & 170554 & 38.7 & 1.0  & 0.38 & 0.38 & 1.0 \\
\Epers{y}(TH=60) & 559          & 216  & 335   & 343 & 174453  & 38.6 & 0.2  & 0.39 & 0.38 & 1.0 \\
\Epers{y}(TH=90) & 241          & 79   & 158   & 162 & 174948  & 32.8 & 0.1  & 0.33 & 0.33 & 1.0 \\ \hline
\end{tabular}\label{tab:geoelectric-events}
\end{table}

\subsection{Forecasting \Ex{} and \Ey{}} \label{res:e}

We first evaluate the LSTMs trained on the geoelectric field in terms of the root-mean-square-error (RMSE) and the Pearson's correlation coefficient (PCC). Comparing the LSTM-E outputs to modelled E, the RMSE values are $126$ mV/km and $111$ mV/km for the absolute value of \Ex{} and \Ey{}, while the PCC values are $0.60$ and $0.61$. Once the sign of E has been included, the RMSE rises to $261$ mV/km and $287$ mV/km, while PCC drops to $0.48$ and $0.32$, so we see that the model's inability to forecast the field direction reliably decreases the accuracy when also considering the field direction.

\textbf{Table \ref{tab:geoelectric-events}} presents an event-based analysis of the LSTM-E results. Multiple thresholds (TH) defining events were considered, and these are listed by the variable "TH" in each line (at $30$, $60$, and $90$ mV/km, representing minor, moderate and strong geomagnetic activity). We see that the skill decreases as the threshold increases (decreasing probability of detection POD and TSS), and that the LSTMs tend towards over-predicting (BS $> 1$). (The bias for the PERS models is always $\sim 1$ because the time series being compared are only shifted in time and therefore almost statistically equivalent.) There are always a large number of false positives, although this remains a small fraction of the number of total data points. The LSTM-E models generally outperform the PERS approach, although the Heidke Skill Scores are occasionally smaller in the LSTMs, which implies a worse balance between false positives and true positives. As in the point-to-point values, the \Ex{} component tends to be predicted better than the \Ey{} component. By evaluating the ROC and DET curves in Fig. \ref{fig:roc-det-all} (a-b), we see that the LSTM-E models outperforms persistence at all thresholds.

We also conducted a comparison with the results from \citeA{Honkonen2018} and \citeA{Lotz2017}, where possible. While the time development of the geoelectric field appears better in the modelling approach in \citeA{Honkonen2018}, the magnitudes are not matched as well. An event-based analysis could not be carried out in their case due to the short time series and lack of larger events, but the RMSE and PCC values for \Ex{} and \Ey{} (reduced to a 15-min sampling rate) come out as $10.5$ mV/km and $97.8$ mV/km and $0.62$ and $0.25$, respectively, which is better in the case of \Ex{} but worse in the case of \Ey{}. Comparing to \citeA{Lotz2017}, we see similar correlations for the geoelectric field components. They found a slightly higher correlation (averaged over three stations and two storms, $0.71$ for \Ex{} and $0.53$ for \Ey{}), although they predicted the maximum value for a longer time span (90 mins), making their approach closer to a nowcast than a forecast. \add{The higher RMSE values seen in our study in part derive from the slightly higher levels of daily variation that is forecast even when the field is extremely quiet. }Again, in both studies used as comparison we see the northward component of the geoelectric field was predicted better than the eastward component.

\begin{table}[h!]
\caption{Metrics from an event-based analysis of different model applied to the years 2017, 2019 and 2020 in a retrospective real-time mode with the model being run at 15-minute intervals. \GICpredE{1} is the result from the models trained to predict the geoelectric field (LSTM-E), while \GICpred{1} is the result from the LSTM-GIC. PERS is a persistence model assuming the target (GIC) repeats itself. The first four columns provide the values for the confusion matrix (where TP, FP, TN and FN are the true positives (hits), false positives, true negatives (misses) and false negatives), the probability of detection (POD), probability of false detection (POFD), Heidke Skill Score (HSS), True Skill Score (TSS), and bias (BS). The variable TH in brackets is the event threshold used to define events and compute the metrics.``undef." refers to the HSS and TSS at TP=0, which are undefined.}
\begin{tabular}{l|llllllllll}
LSTM-E Model       & N$_{events,obs}$ & TP  & FP   & FN  & TN    & POD  & POFD & HSS   & TSS   & BS  \\ \hline
\GICpredE{1}(TH=2) & 432          & 124 & 1060 & 308 & 103697 & 28.7 & 1.0  & 0.15  & 0.28  & 2.7 \\
\GICpredE{1}(TH=4) & 24           & 2   & 57   & 22 & 105108  & 8.3  & 0.1  & 0.05  & 0.08  & 2.5 \\ \hline
\GICpredE{5}(TH=2) & 307          & 159 & 681  & 148 & 80649  & 51.8 & 0.8  & 0.27  & 0.51  & 2.7 \\
\GICpredE{5}(TH=4) & 43           & 6   & 13    & 37 & 81581  & 14.0 & 0.0  & 0.19  & 0.14  & 0.4 \\ \hline
LSTM-GIC Model     & N$_{events,obs}$ & TP  & FP  & FN  & TN     & POD  & POFD & HSS   & TSS   & BS  \\ \hline
\GICpred{1}(TH=2)  & 432          & 239 & 1886 & 193 & 102871 & 55.3 & 1.8  & 0.18  & 0.54  & 4.9 \\
\GICpred{1}(TH=4)  & 24           & 3   & 26   & 21 & 105139  & 12.5 & 0.0  & 0.11  & 0.12  & 1.2 \\ \hline
\GICpred{5}(TH=2)  & 307          & 172 & 1403  & 135 & 79927  & 56.0 & 1.7  & 0.18  & 0.54  & 5.1 \\
\GICpred{5}(TH=4)  & 43           & 7   & 16   & 36 & 81578   & 16.3 & 0.0  & 0.21  & 0.16  & 0.5 \\ \hline
PERS Model         & N$_{events,obs}$ & TP  & FP   & FN & TN     & POD  & POFD & HSS   & TSS   & BS  \\ \hline
\GICpers{1}(TH=2)  & 432          & 50  & 375  & 382 & 104382 & 11.6 & 0.4  & 0.11  & 0.11  & 1.0 \\
\GICpers{1}(TH=4)  & 24           & 0   & 26   & 24 & 105139  & 0.0  & 0.0  & undef. & undef. & 1.1 \\ \hline
\GICpers{5}(TH=2)  & 307          & 61  & 237  & 246 & 81093  & 19.9 & 0.3  & 0.20  & 0.20  & 1.0 \\
\GICpers{5}(TH=4)  & 43           & 7   & 38   & 36 & 81556   & 16.3 & 0.0  & 0.16  & 0.16  & 1.0 \\
\end{tabular}\label{tab:gic-events}
\end{table}

\subsection{Forecasting GICs} \label{res:gics}

The same results are presented for GICs as for the geoelectric field components in the last section. In the event-based analysis, the thresholds were set at $2$, $4$ and $6$~A, which are roughly equivalent to the thresholds used for the electric field. \textbf{Table~\ref{tab:gic-events}} shows the results of this analysis applied to the test data set years 2017, 2019 and 2020, while \textbf{Fig.~\ref{fig:roc-det-all}} depicts the ROC and DET curves for the model output versus measured GICs. A comparison between the LSTM-GIC output and the modelled GICs the model was trained on shows similar levels of accuracy as in LSTM-E to the geoelectric field.

We first look at the results for GICs calculated from the geoelectric field components predicted using the LSTM-E models. Note that while the last section mainly looked at the absolute value of the geoelectric fields, in the calculation of GICs the direction of the geoelectric field is also included, making this an additional error factor if the sign is not predicted accurately. Once the GICs have been calculated using the results from the LSTM-E models and Eq.~\ref{eq:gicfit}, the absolute value is taken for the rest of the analysis. 

As can be seen in \textbf{Table \ref{tab:gic-events}}, the GICs derived from the LSTM-E models see a considerable drop in accuracy in comparison to the results for E alone in Table~\ref{tab:geoelectric-events}. Although there were quite reasonable values for POD predicting E, the POD for GICs at the mid-range threshold ($60$~mV/km or $4$~A) drops from around 50\% in both components of E to 8\% and 16\% in substation SS1 and SS5. Evaluating the skill of the model for GICs at high levels is difficult because there are so few events exceeding even a minimal value of $6$~A. None of these events (2 at SS1, 12 at SS5 over the three years of data) were predicted using any approach.

\begin{figure}[t]
    \centering
    \includegraphics[width=1.0\textwidth]{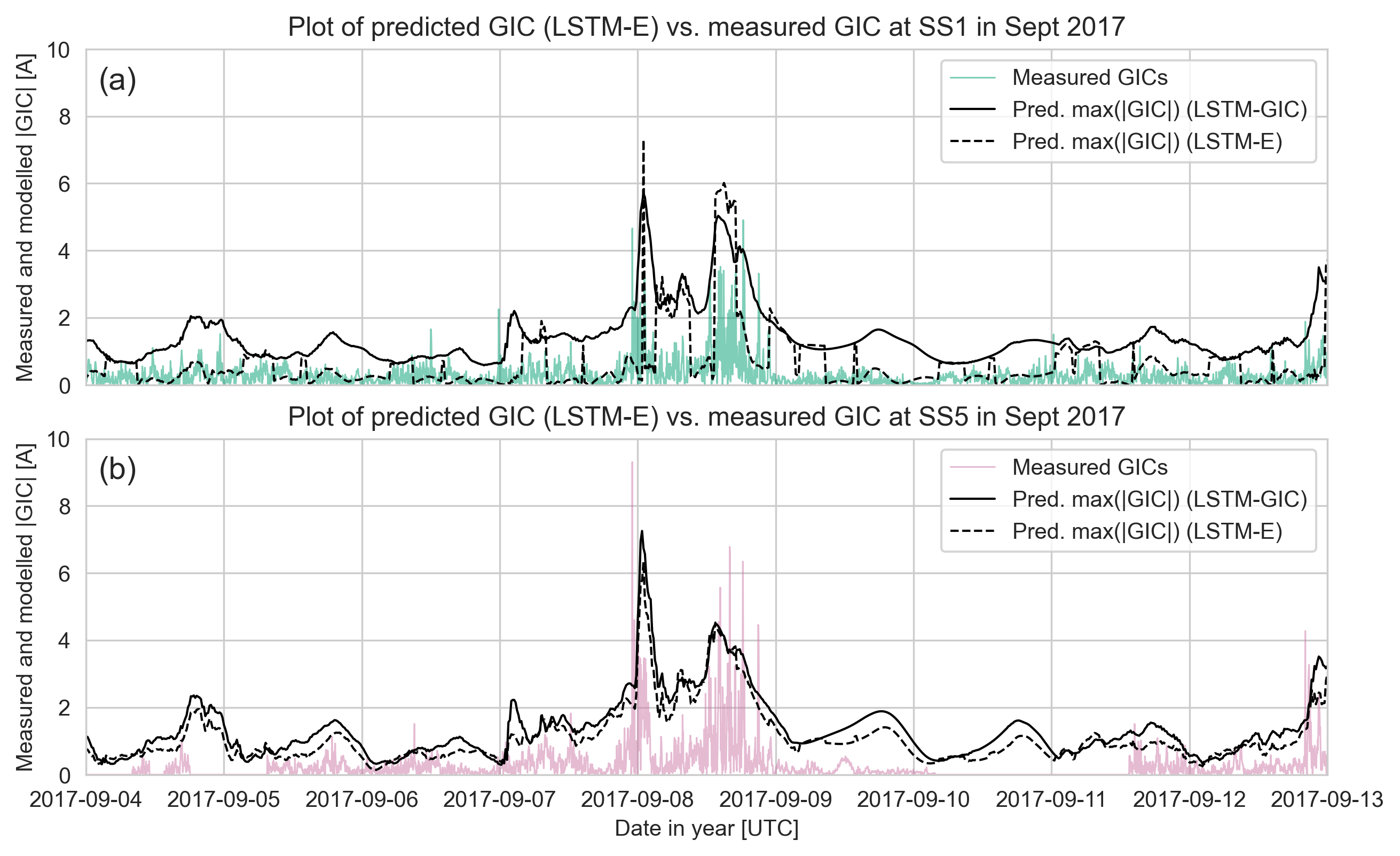}
    \caption{The LSTM-E (dashed line) and LSTM-GIC (solid line) applied to forecasts in an experimental real-time mode and compared to measurements of GICs (coloured lines) at two stations in Austria. The upper panel (a) shows results for SS1 near Vienna, while the lower panel (b) shows results for SS5 near Salzburg (with some data gaps). Although not plotted here, the maximum GIC value computed from the measurements is at the same cadence of 15 minutes to compare to the model forecasts.}
    \label{fig:gic-forecast}
\end{figure}

In comparing the GIC predictions from the two methods (LSTM-E and LSTM-GIC), we see that the LSTM-GIC seems to perform better but the results are station-specific. The LSTM-GIC performs much better than the LSTM-E at SS1 (e.g. a POD of 55\% rather than 29\% and higher HSS and TSS values at a threshold of $2$ A) and at a similar level at SS5. This is also reflected in a model evaluation using point-to-point metrics. The RMSE values for SS1 and SS5 predicted using LSTM-E are $0.49$ A and $0.59$ A, while the PCC is $0.35$ and $0.67$. For GICs predicted using LSTM-GIC, the RMSE values are $0.67$ A and $0.78$ A (i.e. slightly worse than LSTM-E), but the PCC is $0.56$ and $0.64$. The accuracy between the two approaches is roughly equivalent for SS5, but using LSTM-GIC rather than LSTM-E is a definite improvement for SS1 observations. Some of the reason for this can be seen in \textbf{Fig. \ref{fig:gic-forecast}}. In SS1, the jumps in values computed from LSTM-E result from changes in the sign of the geoelectric field components, which then cancel each other out and lead to a GIC of zero. (Conversely, ignoring the sign from LSTM-E and taking the absolute values to calculate the GICs in SS1 results in higher correlation and POD but a far larger number of false positives, leaving this as another possibility.) In the best cases, the GIC forecasts only reach a POD of 16\% for GICs above a threshold of $4$ A, highlighting the difficulty in correctly predicting larger values.

In the ROC and DET curves in \textbf{Figure~\ref{fig:roc-det-all}} panels (c-d) for GICs from LSTM-E and (e-f) from LSTM-GIC, we also see some of the weak forecasting ability for SS1 primarily represents the LSTM behaviour at low values (GICs $< 1$~A). At SS1, there is a mostly continuous level of noise around $1$~A, and the model does not predict the noise while the persistence model captures it clearly. This is an example of the weakness of ROC curves, where in this case only the lower left corner (showing values greater than $1$~A) is of interest to us. 

\textbf{Figure \ref{fig:gic-forecast}} shows the forecast that would have been produced by the model (solid and black dashed lines) against measurements (coloured lines) during the September 2017 storm. The models, particularly the LSTM-GIC approach, do a reasonable job at predicting magnitudes, although the LSTM-E struggles to predict the direction, which is also important for accurate GIC prediction. The storm and the active periods are clearly captured by the forecast, and daily variations from the Sq current are forecasted otherwise. Note that the delayed rise in the forecast of the first peak of the storm does not indicate a timing error. A cross-correlation of the model output shows at maximum an offset in time of 10~minutes and the delay in the figure is simply a feature unique to this storm. While the exact time development of the storm is not captured well, the general scales of GICs are matched well, as is the differentiation between quiet and active times.

In summary, prediction of geoelectric field magnitudes can be achieved with reasonable accuracy (POD of at least 35\% even at the highest event threshold), but the prediction of elevated levels of GICs proves difficult with any approach used. The LSTMs usually outperform the persistence models, except in the bias, where the persistence model has the benefit of being statistically equivalent to the data it is being compared to. The persistence model also generally has a lower POFD and higher HSS value at low thresholds (e.g. TH=$30$ V/km for LSTM-E) because quiet periods tend to persist over time. The LSTMs, however, outperform persistence at the higher thresholds, which are more important for forecasting purposes.

\section{Discussion} \label{sec:discussion}

We have attempted to forecast GICs from solar wind data using LSTMs with two different approaches. We now look at some of the reasons behind the particular difficulty in forecasting GICs. 

Some of the low skill seen when comparing predictions to GIC measurements is down to \change{three}{four} reasons, mostly related to our GIC data: firstly, there is noise in the GIC observations, particularly at SS1, which has a consistent level of $1$~A noise during the day - this is not predicted by the model. Secondly, GIC observations until 2021 had a maximum cutoff point of $3.4$~A in the positive direction, removing some peaks from our event list, and these have not been accounted for.\add{ Thirdly, the model struggles to predict the direction of the geoelectric field values, which are likely driven by smaller-scale ionospheric currents} \cite{Dimmock2020}. \change{Thirdly}{Fourthly}, as noted in Sec.~\ref{sec:gics}, the peaks of observed GICs are often underestimated by geophysical modelling, meaning peaks in the GIC measurements after the cut-off level was removed were often much larger than modelled. This is a problem related to the geoelectric field modelling that may affect the LSTM's ability to learn the problem due to insufficient accuracy in the field modelling. While minute cadence data does capture most of the variability in the GICs, the lack of higher frequency content appears to the primary cause of underestimated peaks, a problem discussed before in \citeA{Grawe2018} and recently for the specific problem of GIC estimates in \citeA{Beggan2021}. As such, it is not surprising that the LSTMs tend to underestimate the actual GICs, and a correction would have to be applied to the target data to account for this.

Outside of the data-specific problems, there are also some timing errors, meaning some peaks arrived slightly later or earlier than they were observed, and as such are not logged as correct predictions even though an event threshold was crossed.

In an application of the model in operations, one caveat is that the maximum possible forecast is $200$~mV/km due to a self-imposed limit to improve the model's ability to learn. We assume that in practise, this would be negligible because all values above a certain level (e.g. $100$~mV/km) would be of interest, regardless of how large they become. As also discussed in \citeA{Wintoft2016}, the scale of geomagnetic variations during extreme events can theoretically become so large that it is effectively unbounded for the purpose of this discussion. In the future, this $200$~mV/km limit could be improved on by training a model specifically for large value forecasting, which can be switched to if the original model forecasts $E > 150$~mV/km.

In an ideal case, a forecasting model would be developed while taking a cost-loss analysis \cite{Murphy1977} such as that used in a space weather context in \citeA{Owens2014} into consideration. In the case of network protection, this is a very complex scenario due to the varying impacts and costs associated with transformer damage or power grid outage, many of which are currently nearly impossible to estimate. This is something that can hopefully be developed further as studies into GIC risk progress \cite{Eastwood2018}.

Another, more general problem in forecasting any measure of ground geomagnetic activity from solar wind measurements without further input from \change[]{magnetospheric modelling}{the magnetosphere-ionosphere system} is that not all geomagnetic variations are driven by the solar wind directly \cite<see e.g.>{Kamide1998, Eastwood2015}. \change[]{Many variations will result from reconnection in the magnetotail}{Many of the ground variations, particularly at shorter timescales} \cite{Alberti2017}\add{, are not directly driven by the solar wind but are instead the consequence of other processes being triggered. These can include complex magnetospheric dynamics such as reconnection in the magnetotail, as well as random, chaotic processes.} \change[]{and would not be relatable}{Such processes can not be related in detail} through our model, which is essentially a coupling function from the solar wind at the bow shock to the geoelectric field in Austria\add[]{. Some of the dynamics will be represented to some degree, but it is difficult to ascertain exactly which in a black-box machine learning model}.\add{ A further difficulty in improving predictions lies in the fact that GICs can only be calculated accurately with knowledge of magnetic field variations at timescales of seconds} \cite{Grawe2018}\add{, ideally, and the LSTM must make approximations of what kind of variations are expected due to the conditions rather than deriving the variations precisely.} Although the machine learning approach described here works at a basic level and could be more promising than forecasts of $dB/dt$ alone, \add[]{to create a model that can also account for complex magnetospheric processes }it would need to be coupled with\add{ either data from space-borne monitors observing the Earth's magnetosphere,} more complex physical models of magnetospheric behaviour \add{(developing a so-called grey-box model as recommended in }\citeA{Camporeale2019}\add{, for example), or both}\remove{ to escape this limitation}.

\add{The calculations and measurements of the GICs shown in this study are for a specific grid configuration, even though the power grid is continually being upgraded and changed. These changes can have large effects on individual GIC scales over long time ranges. The results shown in Table} \ref{tab:active-days} \add{extend far into the past, for which we do not have a detailed history of grid changes, so the values listed could have been much smaller or much larger depending on how the grid was set up. For the LSTM predictions, we have conducted our analysis with the comparison to measurements over a considerably shorter time range of a few years, where the grid has not changed to any great degree, but the predictions may not be valid in the future for a different grid configuration. In this case, a new fit would need to be found for Eq.} \ref{eq:gicfit}, \add{and either the LSTM-GIC model would need to be retrained on the updated GIC data, or the GIC values could be calculated anew from the otherwise unchanged LSTM-E output.}

Our aim was to develop a model that can provide useful forecasts for power grid operators by providing estimates of the scales of GICs. The difference between this and former studies such as \citeA{Lotz2017a} and \citeA{Honkonen2018}, who also predicted ground geoelectric fields from solar wind data, is that we have approached the problem with a new tool (a recurrent neural network) and have been able to forecast GICs directly along with the geoelectric field, with the results compared to measured GICs. We have had some success, particularly with forecasting the geoelectric field, and have tried forecasting substation-specific GICs for the first time, but there are still many problems to be addressed to turn this method into a useful forecast.

\section{Summary} \label{sec:summary}

We have developed a machine learning approach to forecast GICs in Austria. Using data from the past 26 years and the 2003 Halloween storm as a case study, we argued that forecasts of $dB/dt$ alone, which have been the focus of most past studies, are not sufficient to make actionable GIC forecasts. 

From this initial analysis, we set out to forecast maximum expected GICs (over a forty minute window) either directly for specific substations in the power grid or more generally from forecasts of the regional geoelectric field components. From a small set of initial machine learning approaches, an LSTM (recurrent neural network) with an Attention mechanism showed the most promise in forecasting skill and this was developed into a more complex approach.

A selection of models were trained on 21 years of geoelectric field values modelled from geomagnetic variations at the geomagnetic observatory in F\"urstenfeldbruck close to Austria. In the first method, two recurrent neural networks or LSTMs were trained to predict the northward and eastward modelled geoelectric field components and compute the specific substation GICs using a linear equation. In the second method, an LSTM was trained to predict modelled GICs at two substations, which we know correlate very well with the measurements. Five years of data were reserved for testing and evaluating the model. The results were compared to DC measurements at two substations in the Austrian power grid.

The LSTM model worked with reasonable success when predicting the geoelectric field modelled from geomagnetic variations, although translating this success into good GIC forecasts proved difficult. It was possible, however, to outperform a model that simply takes the last observed GICs to forecast future values.

We conclude that forecasting the GICs observed in the power grid from solar wind data measured at L1 is a difficult task, even when the forecasting model does a reasonable job of forecasting the geoelectric field components or modelled GIC. There are many ways to improve the modelling in the future, including using higher-resolution magnetic field measurements (or applying a correction to the modelled geoelectric field before training) to more accurately estimate the peak geoelectric field and GIC values, and by including information on the development of the magnetosphere during storm times.

Although this study has looked specifically at a mid-latitude region, where geomagnetic variations and GICs are not as large as those seen in higher latitude regions such as Scandinavia, we have been able to compare model output directly to measurements and expect that the conclusions drawn will also be valid for other regions with GICs at different scales.

A lower-resolution version of the LSTM-E model will be coupled with the PREDSTORM solar wind forecast \cite{Bailey2020}, which provides forecasts of the ambient solar wind a few days in advance, based on either a recurrence model or data from a spacecraft east of the Sun-Earth line such as STEREO or a future mission to the Lagrange 5 point. We also plan in the future to integrate methods on solar wind $B_z$ forecasting \cite{reiss2021bz} or CME flux rope modelling \cite{Weiss2021} to advance our capabilities in GIC forecasting for any type of solar wind structures.

\section{Data Availability} \label{sec:datasources}

\begin{itemize}
    \item INTERMAGNET data for FUR and WIC:\\ \url{https://intermagnet.org/data-donnee/download-eng.php}
    \item OMNI data: \url{https://spdf.gsfc.nasa.gov/pub/data/omni/high_res_omni/}
    \item Open source code for this work (in Python 3 and Jupyter Notebook form):\\ \url{https://doi.org/10.5281/zenodo.5704715}
    \item Exact details on the LSTM structure and hyperparameters used for training can be found in the supporting information for this study.
    \item \add{A subset of the data set used to derive the results, namely the the GIC observations and model forecasts used to produce Figure} \ref{fig:gic-forecast}, \add{have also been included in the supporting information and saved in an online repository: \newline \url{https://doi.org/10.6084/m9.figshare.19102772.v1}}
\end{itemize}


%
%
%
%
%
%
%
%

\acknowledgments
The results presented in this paper rely on the data collected at the Conrad Observatory (ZAMG) in Austria and at Fürstenfeldbruck (LMU), Germany. We thank the Ludwig-Maximilians-Universität München for supporting its operation and INTERMAGNET for promoting high standards of magnetic observatory practice (\url{www.intermagnet.org}). The data used in this study is publicly available (with the exception of the measurements of GICs in Austria), and details on where to find the data can be found in Section \ref{sec:datasources}.\add[]{ An excerpt of GIC measurements and model output has been provided in the Supporting Information, and we thank both the TU Graz and Austrian Power Grid for supporting these measurements.} We thank Ilja Honkonen for providing the results from his 2018 study for comparison. R.L.B., C.M., M.A.R., and A.J.W.~thank the Austrian Science Fund (FWF) for research funding from projects P31659-N27 and P31521-N27. C.D.B. was funded under UK Natural Environment Research Council Grant NE/P017231/1 ``Space Weather Impact on Ground-based Systems (SWIGS)''. \add{We thank the two anonymous reviewers for their careful readings of this manuscript and helpful suggestions.}


\bibliography{bib}

\end{document}